# How to design multi-target drugs: Target search options in cellular networks

**Tamás Korcsmáros, Máté S Szalay, Csaba Böde, István A Kovács & Péter Csermely**[†]
[†]Department of Medical Chemistry, Semmelweis University, P O Box 260., H-1444 Budapest 8, Hungary and Predinet Ltd., Dongo str. 8., H-1149 Budapest, Hungary

Despite improved rational drug design and a remarkable progress in genomic, proteomic and high-throughput screening methods, the number of novel, single-target drugs fell much behind expectations during the past decade. Multi-target drugs multiply the number of pharmacologically relevant target molecules by introducing a set of indirect, network-dependent effects. Parallel with this the low-affinity binding of multi-target drugs eases the constraints of druggability, and significantly increases the size of the druggable proteome. These effects tremendously expand the number of potential drug targets, and will introduce novel classes of multi-target drugs with smaller side effects and toxicity. Here we review the recent progress in this field, compare possible network attack strategies, and propose several methods to find target-sets for multi-target drugs.

**1. Introduction: emergence and rationale of the multi-drug concept**
Recent drug development strategies were based on the emergence of potential targets in genomic and proteomic studies. Therefore, the currently followed drug-development paradigm can be summarized as to: (a) find a target of clinical relevance; (b) identify the 'best-binder druggable molecule' by high-throughput screening of large combinatorial libraries and/or by rational drug design based on the three-dimensional structure of the target; (c) provide a set of proof-of-principle experiments; and (d) develop a technology platform leading to clinical applications. However, despite all the considerable drug-development efforts undertaken, the number of successful drugs and novel targets fell significantly behind the expectations during past decades [1-3].

A number of novel strategies have been developed to overcome the target-shortage, and to add novel classes of drugs to development pipelines. Many of these drug development directions aim to influence multiple targets in a parallel fashion. One of the most widespread multiple target approaches, combination therapy is increasingly used to treat many types of diseases, such as AIDS, atherosclerosis, cancer and depression. As one of newly developed combination therapies, 'multi-target lead discovery' is a promising tool for the identification of unexpectedly novel effects of drug combinations [4-8]. Recently, initial steps have been taken to develop aptamer combinations against complex sets of targets [9].

Multiple target strategies have only been re-discovered by drug-developers. Snake and spider venoms are both multi-component systems, and plants also developed a combinative-strategy to defend themselves against pathogens. Additionally, traditional medicaments and remedies often contain multi-component extracts of natural products [10,11]. All these examples show that multiple target strategies have benefits, which were utilized as medicaments by our ancestors several thousand years ago, and withstood million-year evolutionary selection.

Agents aiming only a single target ('single-hits') might not always affect complex systems in the desired way even if they completely change the behaviour of their immediate target. Single targets might have 'back-up' systems that are sometimes different enough not to respond to the same drug. Moreover, many cellular networks are robust and prevent major changes in their outputs despite dramatic changes in their constituents [12-15]. These considerations are independent of whether or not the pharmacological agent inhibits or activates its target.

## 2. Examples for multi-target strategies

Several efficient drugs, such as salicylate, non-steroidal anti-inflammatory drugs (NSAIDs), metformin, anti-depressants, anti-neurodegenerative agents, multi-target kinase inhibitors (such as Gleevec™, or the inhibitors of the kinase-maturating molecular chaperone, Hsp90) affect many targets simultaneously [7,16-24]. Multi-target antibodies (in forms of diabodies, triabodies, tetrabodies and recombinant polyclonal antibodies) are increasingly used in cancer therapy to delay the development of resistance [25,26].

We may find a large number of currently used terms to describe ligands that have multiple activities: the words balanced, binary, bivalent, dimeric, dual, mixed or triple are all used in combination with various suffixes, such as agonist, antagonist, blocker, conjugate, inhibitor or ligand. Various pharmacophores may have an increasing overlap, which actually gives an almost continuous spectrum starting from conjugates (or cleavable conjugates, which are actually a novel chemical form of combinational therapies) through overlapping pharmacophores till the highly integrated multi-target drugs (**Figure 1**; [7,19]).

Multi-target drugs offer a magnification of the 'sweet spot' of drug discovery [1], meaning the overlap between pathways, which are interesting from the pharmacological point of view, and the hits of chemical proteomics, which represent those proteins, which can interact with druggable molecules (meaning small, hydrophobic molecules with a good bioavailability). The 'sweet spot' represents those few hundred proteins, which are both parts of interesting pathways and are druggable [1]. The option to allow indirect effects via network-contacts of multi-target drugs expands the first circle, since the number of those proteins, which are indirectly related to existing targets of pharmacologically important pathways, is by magnitudes greater than the number of the targets themselves. On the other hand, the low-affinity binding of multi-target drugs enlarges the second circle, since it eases the constraints of druggability. (Small, hydrophobic molecules bind to only a small subset of proteins with high affinity. However, the very same molecules interact with ten or even hundred times more proteins with increasingly lower and lower affinity. Low affinity binding here describes interactions with dissociation constants in the higher micromolar or even close to the milimolar range. Low affinity binding also implies a more transient interaction, where the off-rate is comparable or higher than the on-rate.) As a result of these combined effects the 'sweet spot' of drug discovery may easily become a wide 'candy-field' (**Figure 2**).

## 3. Cellular networks: drug target maps

Cellular networks help us to understand the complexity of the cell. In the network concept the complex system is perceived as a set of interacting elements, which are bound together by links. Links usually have a weight, which characterizes their strength (affinity, or propensity). Links may also be directed links, when one of the elements has a larger influence to the other than vice versa. In cellular networks the interacting molecules are considered as the elements, and their interactions form the weighted, but not necessarily directed links of the respective structural network. Alternatively, we may also envision directed links as representations of signalling or metabolic processes of the functional networks in the cell (**Table 1.** [27-29]). Cellular networks often form small worlds, where two elements of the network are separated by only a few other elements. Networks of our cells contain hubs, i.e. elements, which have a large number of neighbours. These networks can be dissected to overlapping modules, which form hierarchical communities [30-32]. However, this summary of the major features of cellular networks is largely a generalization, and needs to be validated through critical scrutiny of the datasets, sampling procedures and methods of data analysis at each network examined [33,34].

Cellular networks offer a lot of possibilities to point out their key elements as potential drug targets. As an example of these possibilities, signalling networks have interdigitated pathways and multiple layers of cross-talks [35]. Special signalling elements, such as the PI-3-kinase, the Akt-kinase, or the insulin-receptor substrate-family have been called 'critical nodes'. These 'critical nodes' have multiple isoforms, and are important junctions of signalling pathways [36]. Both the bridge-elements of signalling networks providing cross-talks and the 'critical nodes' can be important targets of network-based drug development. Domain-specific target-analysis of protein-protein interaction networks extends the map of physical interactions towards functional understanding. Domain-specific targets offer a larger flexibility and may actually reflect a family of multiple targets due to frequent 're-use' of domain-variants as a result of modular evolution [37,38]. Elements of metabolic and cytoskeletal networks have also been analyzed as drug targets [39,40].

However, current databases of most cellular networks suffer a lot of uncertainties. Protein-protein interaction networks have a large number of false-positive entries, which makes the inclusion and assessment of low-affinity interactions especially difficult. Moreover, current databases mostly give an averaged probability of the particular interaction. This does not take into account, if the two proteins are expressed at the same time, or they are located in the same cellular compartment at the same time. Most databases do not contain the information of the ratio of two interacting proteins in the given status of the particular healthy or 'sick' cell. Literature-derived,

'evidence-based' databases suffer from nomenclature and interpretation problems of the original data. However, recent advances connect protein-protein interaction databases with protein structure data, which make both the validation and prediction of protein-protein interactions more robust [41-43].

We may overcome with the above problem by using curated databases, which contain only the most valid, most accurate information. However, these databases will miss most low-affinity interactions, and we loose around 80% of the available information due to our increased scrutiny. As an alternative approach we keep all information taking into account that our database becomes 'fuzzy' due to the inclusion of potentially false data. In this approach to correct the errors, we need highly integrated methods for network analysis, which are able to build in all the above information and take into account those less, which are in contradiction with most of the others. Advantageously, these integrated analytical tools should be used in a 'zoom-in' fashion, where the user may define the 'integration-level' of her choice. Low resolution network maps can be calculated fast, and by directing the user's mind, show the most important 'take-home messages' of the analysis. Zooming-in to a high-resolution analysis with the same, flexible method, will show the refined details of all available information and gives the user a large number of correct (and false) ideas to think about and to test in experiments either in the primary 'hit areas' of the low-resolution analysis, or the spots of specific interests based on other assumptions.

**4. Multi-target drugs are often low-affinity binders**
Development of a multi-target drug is likely to produce a drug, which is interacting with its target having a lower affinity than a single-target drug because it is unlikely that a small, drug-like molecule can bind to a number of different targets with equally high affinity. However, low-affinity drug binding is apparently not a disadvantage. For example, memantine (a drug used to treat Alzheimer's disease) and other multi-target non-competitive NMDA receptor antagonists show that low-affinity, multi-target drugs might have a lower prevalence and a reduced range of side-effects than high-affinity, single-target drugs [20,44]. The recent suggestion to use unstructured proteins, as a novel and un-explored field of drug-targets uses the beneficial effects of low-affinity, but rather specific binding, which has been shown extremely useful in regulation and signalling [45]. (Here again, low affinity binding denotes interactions with dissociation constants in the higher micromolar or even close to the milimolar range. Low affinity binding also implies a rather transient interaction, where the off-rate is comparable or higher than the on-rate.)

Does low-affinity binding predict a low-efficiency? Not necessarily. A vast majority (>80%) of the cellular protein, signalling and transcriptional networks are in a low-affinity, or transient 'weak linkage' with each other. In metabolic networks, weak links are those reactions, which have a low flux [29,46]. In this paper we use the term 'weak linker' to denote small molecules and drugs that interact with cellular proteins having a low-affinity. Thus, most multi-target drugs are weak linkers. Because most links in cellular networks are weak, a low-affinity multi-target drug might be sufficient to achieve a significant modification. The recent paper of Bruce N. Ames [47] on the potential impact of micro-nutrients on disease-development is a good example of the profound effects emerging from seemingly minor interactions. Low affinity, 'imperfect' binding allows the development of special, cooperative binding-behaviour, which may lead to a switch-type activation setting a threshold for various cellular events, such as DNA-replication [48,49].

**5. Identification of drug targets using the network approach: attack strategies**
Drug design strategies are mostly based on target-driven approaches, where an efficient compound to influence disease-related molecular target is sought. The network approach examines the effects of drugs in the context of cellular networks. In this model a drug-induced inhibition of a single target means that the interactions around a given target are eliminated, whereas partial inhibition can be modelled as a partial knockout of the interactions of the target.

Cellular networks are usually damaged by random failures, such as the oxidative damage of free radicals, the indirect effect of somatic mutations as well as the complex phenomenon of ageing [50]. Opposed to this, drug-driven network attacks are targeted to find the most efficient way to influence network behaviour. Several classes of drugs such as antibiotics, fungicides, anti-cancer drugs as well as numerous chemical compounds such as pesticides are designed to destroy the normal function of cellular networks.

Networks have a number of vulnerable points, and, therefore, can be attacked in many ways (**Figure 3**). The first major insight identifying a set of 'weak points' in may natural networks came from the work of Albert-Laszlo Barabasi and co-workers [30,51], who uncovered that many real-world networks (including cellular networks) have a more-less scale-free degree distribution, which means (in simple terms) that these networks have hubs (e.g. nodes with a much greater number of neighbours than average). If hubs are selectively attacked, the information-transfer in most scale-free networks soon becomes significantly hindered. In other words: hubs are central points of networks. On the contrary, scale-free networks are highly resistant against random damage. These two features can be summarized as a 'robust-but-fragile' nature of scale free networks. As an illustration for the robustness, we may delete 99% of the Internet in random-attack strategy and the continuity of the network still remains, i.e. we could still use the Internet after such an attack, albeit it would be much slower than usual [52]. On the contrary, malicious attacks on hubs may follow a 'greedy' strategy meaning that degrees are continuously re-calculated after each attack and network elements are re-ranked. This greedy strategy is often more powerful, than using the original degrees of network elements throughout the whole process [53].

Hubs are the centres of networks only from the point of local network topology. Another approach for pinpointing central elements of network communication is to find those elements (or links), which are in a centred position not in the local, but in the global topology. Betweenness centrality of a link refers to the number of shortest paths between any two elements of the network across the given link. Betweenness centrality was worked out first in social networks [54] but later it became a preferred centrality measure to assess the presumed effect of targeted attacks on network stability. Inverse geodesic length (also called network efficiency) meaning the sum of the inverse of the shortest paths between network elements is a widely used indicator of network damage after the removal of links or elements [53,55,56]. Alternative measures of damaging element or link removal have also been worked out by Latora and Marchiori [57], which are based on monitoring of the performance of the whole network.

Recent studies take into account the weights and directedness of network links. This is much closer to the real, cellular scenario, where protein-protein interactions are characterized by their affinity and/or prevalence (link weight) as well as direction (e.g. in form of signalling). The removal of the links with the highest weighted centralities is often more devastating to network behaviour than the removal of the most central links based on the un-weighted version of the same network topology [58].

Another recent approach is to take into account 'mesoscopic' centrality network topology measures, which are neither based on local information (such as hubs) or global information (such as betweenness centrality) on network structure. Motter et al. [59] found that the removal of 'long-range' links connecting elements lying in a long distance from each other has a profound impact on small-world networks, however, it fails to affect many scale-free networks. Another complication is caused by multi-layered networks [60], where various communities, modules are organized in a hierarchical fashion. These studies reveal that the utilization of the complex structural information of real world networks needs more sophisticated methods, such as an integrated assessment of link-density and network topology (Kovacs et al., manuscript in preparation).

The network approach gained an increasing ground in helping drug target analysis by now. Flux-balance analysis (or metabolic control analysis) uses a large database of experimental data, and calculates all metabolic rates of the metabolic network assuming that the rates of reactions producing a metabolite must be equal with the rates of reactions, which consume it. Flux-balance analysis of metabolic networks uncovers vulnerable points of parasite or pathological metabolisms providing potential targets for efficient drug action [39,61-63]. Comparison of cellular (transcriptional, signalling, protein-protein interaction, etc.) networks from various genomes helps to identify the function of novel proteins and thus increases the number of potential drug targets [64,65]. Analysis of protein-protein interactions identifies protein contact surfaces as potential sites of drug action [66] and neural networks have long been applied in the methodological and computational help of drug design [67].

Most of the above network-related methods have been used so far to steer target-identification attempts to single targets and a systematic network-based analysis of multi-target drug action is still to come. In our earlier study [55,68] we modelled a multi-target 'attack' on the genetic regulatory networks of the bacterium *Escherichia coli* or the yeast *Saccharomyces cerevisiae*. A comparison of various strategies suggested that multiple but partial attacks on carefully selected targets were almost inevitably more efficient than the knockout of a single target. For example, the largest damage to the *E. coli* regulatory network was reached by removing an element with 72 connections. However, the same damage could be achieved, if 3 to 5 nodes were partially inactivated. Multiple attacks proved to be more efficient than a single attack even if the number of affected interactions remained the same [55,68]. Thus, the reason underlying the efficiency of multi-target attacks was proved to be not trivial even from a theoretical point of view: multi-target attacks were not only better because they affected the network at more sites, they could, especially if distributed in the entire network, perturb complex systems more than concentrated attacks even if the number of targeted interactions remained the same.

## 5. 'Network diseases'

Our initial network-based analysis of the potential efficiency of multi-target drugs [55,68] was based on the topology of bacterial and yeast gene regulatory networks, which may be regarded as an initial model of the multi-target action of antibiotics and fungicides, where network damage corresponds well to the desired drug action. For the analysis of multi-target drugs affecting specific disease models (e.g. anti-hypertensive, anti-psychotic and anti-diabetic drugs), more specific signalling, metabolic and transcriptional network models are needed. As a prelude of this process several complex, multifactorial diseases have already been described as 'network diseases'. Cancer was assessed as a 'systems biology disease' by Hornberg et al. [69]. The complexity of intra- and extra-cellular cancer-specific changes in signalling, gene-regulatory (and, most probably, protein-protein interaction) networks, the profound reorganization of cellular metabolism, the multiple types and interactions of cells involved, and the complexity of all these events at various types and subtypes of malignant transformation indeed, make the name, 'systems biology disease' well deserved for all stages of tumour development.

Network-effects of continuously changing functional neuron-assemblies may provide an explanation of the daily fluctuations in the symptoms of neurodegenerative diseases. This approach may show novel pathways of drug development leading to shorter and cheaper clinical trials, which concentrate on the short-term attenuation of symptom-fluctuations instead of waiting for and monitoring the long-term, and rather elusive benefits of inhibited neurodegeneration [70]. Since increased fluctuations of the efficiency may reflect a general decline in the stability of the overall network – which is related to the reduction of weak links [29,46] – multi-target drugs, in fact, might be an ideal choice to prevent the further functional loss in this sense. The complexity of neuronal networks also led the concept of network-disease in case of depression aiding the design of novel, multi-target anti-depressants [7].

Last, but not least ageing has also been conceived as a 'network disease' [50,71]. The multiple reasons and stages of biological ageing, the increasing variability of symptoms and malfunctions both from one elderly person to the other and from one day to another all call for a network-based analysis of the increasing amount of data collected so far.

## 6. Target-sets of multi-target drugs – the help of networks

How should we find the relevant target-sets of multi-target drugs? In the last decade several experimental and modelling approaches have been developed to identify single targets in a network context [39,61,62,72-75]. Appropriate modifications of these approaches may constitute the first step for zooming-in to a smaller set of potential targets. A high-throughput combinative screen of all possible combinations may be a daunting task and prospect [5]. How could we pin-point those target-combinations, which might be relevant in the clinical setting?

Surprisingly, 'old fashioned' drug development might come back here to help: *in vivo* pharmacology (i.e. whole-animal studies) might become important again [76]. However, for more efficient *in vivo* testing, better animal models are needed. Better animal models can be achieved by 'humanizing' the metabolic and signalling network of test animals.

Can the network approach help us to suggest potential target-sets? The answer is currently not known. However, we have many promising tools to assess the relevance of our present, network-based knowledge on the complexity of the cell in pathological states. We will list these in the "Expert opinion" section.

## 7. Conclusions

Combinatorial therapies recently become one of the most successful drug development strategies. Multi-target drugs can be perceived as siblings of combinatorial therapies, where the different agents (often as many as 5-6 of them) are comprised to a single, integrated chemical entity. Multi-target drugs expand the number of pharmacologically relevant target molecules by introducing a set of indirect, network-dependent effects. Multi-target drugs usually have a low affinity towards their targets. An increasing number of evidence shows that low affinity – especially, if multiplied – does not mean low efficiency. On the contrary, several signalling and regulatory events are actually based on low-affinity interactions. Moreover, low-affinity binding of multi-target drugs eases the constraints of druggability, and significantly increases the size of the druggable proteome. These effects tremendously expand the number of potential drug targets, and will introduce novel classes of multi-target drugs with smaller side effects and toxicity. Cellular networks offer a large number of possibilities, such as hubs and bridges with high betweenness centrality to find target-sets of multi-target drugs. However, the 'fuzzyness' (meaning uncertain and incomplete information) of cellular networks and the data-sets on 'network diseases' (such as cancer, diabetes and neurodegeneration) need a more sophisticated approach.

**8. Expert opinion: Network-based, smart multi-target drugs of the future**

We predict that in 5-10 years multi-target drugs will be much more common than now. The emerging knowledge of recent years strongly suggests that these drugs have a better chance of affecting the complex equilibrium of whole cellular networks than drugs acting on a single target. Target-expansion to indirect targets will lead to the discovery of several novel classes of drugs. Moreover, it is sufficient that these multi-target drugs affect their targets only partially, which multiplies our target choices from the point of druggability.

Low-affinity, multi-target drugs might have another advantage. Weak links have been shown to stabilize complex networks, including macromolecular networks, ecosystems and social networks, buffering the changes after system perturbations. If multi-target, low-affinity drugs inhibit their targets, they change a strong link into a weak link instead of eliminating the link completely. A weak activation also results in a weak link in most of the cases. Thus, multi-target drugs can increase the number of weak links in cellular networks and thus stabilize these networks in addition to having multiple effects. Stabilization of the cells becomes especially important, if we take into account that cells of stressed, sick and ageing organisms are at the 'verge of chaos', showing a much greater instability than their healthy counterparts [29,46,71,77,78]. Thus multi-target drugs may have multiple beneficial effects:

- multi-target drugs can be designed to act on a carefully selected set of primary targets, which help to sum up the action of the drug on key, therapeutically relevant secondary targets;
- *via* the above, 'indirect' approach, and their low affinity binding multi-target drugs may avoid the presently common dual-trap of drug-resistance and toxicity;
- finally, the low affinity, 'weak links' of multi-target drugs have an important side-effect: they stabilize the sick cell, which may be sometimes at least as beneficial as their primary therapeutic effect.

What are the 5-10 years perspectives to find the relevant target-sets of multi-target drugs? Here, our increasing knowledge of cellular networks will certainly play a key role in the future. The increasing complexity of the datasets requires more sophisticated tools to direct our attention to relevant areas of the information-universe. In the last part of our review we list a few ideas for future directions in the development of network analysis tool-kits:

- overlaps of network communities [31], which have a key role in the regulation of complex systems [29] may be efficient guides to restrict the initial target-pool in search of targets-sets for multi-target drugs;
- differential analysis of relevant cellular networks (such as signalling networks, gene-regulatory networks, metabolic networks and protein-protein interaction networks) of healthy and 'sick' cells may provide an even more efficient screen;
- finally, the current development of analytical tools to assess the 'evolution' and dynamism of whole cellular networks will reveal search methods to de-convolute the hidden masterminds of the primary target-sets from the currently known, pharmacologically relevant secondary-targets (Kovacs et al., patent application submitted). A variant of this approach has been called 'reverse-engineering', which deciphers potential targets by the analysis of the effect of a limited set of experimental perturbations [72,79-81]. These approaches open the way to find the multiple-targets, and to design 'alternative target-sets' to mimic the action of existing, successful drugs.

In summary, 10 years from now we will have a multitude of expanded cellular datasets having detailed and differential information on variable pairs of 'healthy' and 'sick' cells. These datasets will have graded information, which enables to build weighted and directed networks. Analytical tools will be developed to assess the complexity of these networks keeping all data and giving a zoom-in picture of any resolution in a computationally accessible, short time. Multi-target drugs will magnify the currently available target field by introducing thousands of secondary targets as well as other thousands of druggable proteins. This will all lead to the discovery of several entirely novel drug classes. Finally we will enjoy the network option of 'fine-tuning' of drug action of multi-target drugs by targeted manipulation of certain elements of the already existing target-sets of multi-target drugs, which may lead to the development of 'personalized medicines' – at a low cost.

**Acknowledgements** Authors would like to thank the three anonymous referees for their helpful comments. Supported in part by funds from the EU (FP6-506850, FP6-016003) and the Hungarian Office of Research and Development (NFKP-1A/056/2004).

**Bibliography**
1. BROWN D, SUPERTI-FURGA G: Rediscovering the sweet spot in drug discovery. *Drug Discov. Today* (2003) **8**:1067-1077.
    • **An important review on target-search options.**
2. OVERINGTON JP, AL-LAZIKANI B, HOPKINS AL: How many drug targets are there? *Nat. Rev. Drug Discov.* (2006) **5**:993-996.
3. SZUROMI P, VINSON V, MARSHALL E: Rethinking drug discovery. *Science* (2004) **303**:1795
4. BORISY AA, ELLIOTT PJ, HURST NW *et al*: Systematic discovery of multicomponent therapeutics. *Proc. Natl. Acad. Sci. USA* (2003) **100**:7977-7982.
    • **A key paper on combination therapy-based drug design.**
5. DANCEY JE, CHEN HX: Strategies for optimizing combinations of molecularly targeted anticancer agents. *Nat. Rev. Drug Discov.* (2006) **5**:649-659.
6. KEITH CT, BORISY AA, STOCKWELL BR: Multicomponent therapeutics for networked systems. *Nat. Rev. Drug Discov.* (2005) **4**:71-78.
7. MILLAN MJ: Multi-target strategies for the improved treatment of depressive states: Conceptual foundations and neuronal substrates, drug discovery and therapeutic application. *Pharmacol. Therap.* (2006) **110**:135-370.
8. ZIMMERMANN GR, LEHAR J, KEITH CT: Multi-target therapeutics: when the whole is greater than the sum of the parts. *Drug Discov. Today* (2007) **12**:34-42.
9. FITTER S, JAMES R: Deconvolution of a complex target using DNA aptamers. *J. Biol. Chem.* (2005) **280**:34193-34201.
10. ROLLINGER JM, LANGER T, STUPPNER H: Strategies for efficient lead structure discovery from natural products. *Curr. Med. Chem.* (2006)**13**:1491-1507.
11. SHAFIKHANI SH, ENGEL J: Pseudomonas aeruginosa type III-secreted toxin ExoT inhibits host-cell division by targeting cytokinesis at multiple steps. *Proc. Natl. Acad. Sci. USA* (2006) **103**:15605-15610.
12. PAPP B, PAL C, HURST LD: Metabolic network analysis of the causes and evolution of enzyme dispensability in yeast. *Nature* (2004) **429**:661-664.
13. PAL C, PAPP B, LERCHER MJ, CSERMELY P, OLIVER SG, HURST LD: Chance and necessity in the evolution of minimal metabolic networks *Nature* (2006) **440**:667-670.
14. KITANO H: A robustness-based approach to systems-oriented drug design. *Nat. Rev. Drug Discov.* (2007) **6**:202-210.
15. SHAROM JR, BELLOWS DS, TYERS M: From large networks to small molecules. *Curr. Op. Chem. Biol.* (2004) **8**:81-90.
16. FRANTZ S: Drug discovery: Playing dirty. *Nature* (2005) **437**:942-943.
17. HUANG S: Rational drug discovery: what can we learn from regulatory networks? *Drug Discov. Today* (2002) **7**:S163-S169.
18. KAELIN WG Jr.: Gleevec: prototype or outlier? *Science STKE* (2004) (225), pe12
19. MORPHY R, KAY C, RANKOVIC Z: From magic bullets to designed multiple ligands. *Drug Discovery Today* (2004) **9**:641-651.
    • **A structural analysis of multi-target drug design.**
20. ROGAWSKI MA: Low affinity channel blocking (uncompetitive) NMDA receptor antagonists as therapeutic agents – towards an understanding of their favorable tolerability. *Amino Acids* (2000) **19**:133-149.
21. SATHORNSUMETEE S, VREDENBURGH KA, LATTIMORE KP, RICH JN: Malignant glioma drug discovery – targeting protein kinases. *Expert. Opin. Drug Discov.* (2007) **2**:1-17.
22. SOTI C, NAGY E, GIRICZ Z, VIGH L, CSERMELY P, FERDINANDY P: Heat shock proteins as emerging therapeutic targets. *Br. J. Pharmacol.* (2005) **146**:769-780.
23. SREEDHAR AS, SOTI C, CSERMELY P: Inhibition of Hsp90: a new strategy for inhibiting protein kinases. *Biochim. Biophys. Acta* (2004) **1697**:233-242.
24. YOUDIM MBH, BUCCAFUSCO JJ: Multi-functional drugs for various CNS targets in the treatment of neurodegenerative disorders. *Trends Pharmacol. Sci.* (2005) **26**:27-36.
25. SHARON J, LIEBMAN MA, WILLIAMS BR: Recombinant polyclonal antibodies for cancer therapy. *J. Cell Biochem.* (2005) **96**:305-313.
26. TODOROVSKA A, ROOVERS RC, DOLEZAL O, KORTT AA, HOOGENBOOM HR, HUDSON PJ: Design and application of diabodies, triabodies and tetrabodies for cancer targeting. *J. Immunol. Methods* (2001) **248**:47-66.
27. BARABASI A-L, OLTVAI ZN: Network biology: understanding the cell's functional organization. *Nat. Rev. Genet.* (2004) **5**:101-113.
    • **A comprehensive and brief summary of the properties of biological networks.**
28. BOCCALETTI S, LATORA V, MORENO Y, CHAVEZ M, HWANG D-U: Complex networks: structure and dynamics. *Physics Rep.* (2006) **424**:175-308.
29. CSERMELY P: *Weak links: a universal key for network diversity and stability*. Springer Verlag, Heidelberg (2006).
    •**A comprehensive and easily understandable summary of network science from a biological perspective with more than 800 original references.**
30. BARABASI A-L, ALBERT R: Emergence of scaling in random networks. *Science* (1999) **286**:509-512.
    • **A key paper on the generality of scale-free degree distribution of networks.**
31. PALLA G, DERENYI I, FARKAS T, VICSEK T: Uncovering the overlapping community structure of complex networks in nature and society. *Nature* (2005) **435**:814-818.
    • **The paper describes a method to find overlapping network modules.**

32. WATTS DJ, STOGATZ SH: Collective dynamics of 'small-world' networks. *Nature* (1998) **393**:440-442.
    • **A key paper on the generality of the small-world character of networks.**
33. ARITA M: The metabolic world of Escherichia coli is not small. *Proc. Natl. Acad. Sci. USA* (2004) **101**:1543-1547.
34. TANAKA R, YI TM, DOYLE J: Some protein interaction data do not exhibit power law statistics. *FEBS Lett.* (2005) **579**:5140-5144.
35. CHAUDHURI A, CHANT J: Protein-interaction mapping in search of effective drug targets. *Bioessays* (2005) **27**:958-969.
36. TANIGUCHI CM, EMANUELLI B, KAHN CR: Critical nodes in signalling pathways: insights into insulin action. *Nat. Rev. Mol. Cell Biol.* (2006) **7**:85-96.
37. SANTONICO E, CASTAGNOLI L, CESARENI G: Methods to reveal domain networks. *Drug Discov. Today* (2005) **10**:1111-1117.
38. WILSON D, MADERA M, VOGEL C, CHOTHIA C, GOUGH J: The SUPERFAMILY database in 2007: families and functions. *Nucleic Acids Res.* (2007) **35**:D308-D313.
39. KELL DB: Systems biology, metabolic modelling and metabolomics in drug discovery and development. *Drug Discov. Today* (2006) **11**:1085-1092.
40. MICHAELIS ML, SEYB KI, ANSAR S: Cytoskeletal integrity as a drug target. *Curr. Alzheimer Res.* (2005) **2**:227-229.
41. VON MERING C, KRAUSE R, SNEL B *et al.*: Comparative assessment of large-scale data sets of protein-protein interactions. *Nature* (2002) **417**:399-403.
42. SHEN J, ZHANG J, LUO X *et al.*: Predicting protein-protein interactions based only on sequences information. *Proc. Natl. Acad. Sci. USA* (2007) **104**:4337-4341.
43. KIM PM, LU LJ, XIA Y, GERSTEIN MB: Relating three-dimensional structures to protein networks provides evolutionary insights. *Science* (2006) **314**:1938-1941.
44. LIPTON SA: Turning down, but not off. Neuroprotection requires a paradigm shift in drug development. *Nature* (2004) **428**:473.
45. CHENG Y, LEGALL T, OLDFIELD CJ *et al.*: Rational drug design via intrinsically disordered protein. *Trends Biotechnol.* (2006) **24**:435-442.
46. CSERMELY P: Strong links are important, but weak links stabilize them. *Trends Biochem. Sci.* (2004) **29**:331-334.
47. AMES BN: Low micronutrient intake may accelerate the degenerative diseases of aging through allocation of scarce micronutrients by triage. *Proc. Natl. Acad. Sci. USA* (2006) **103**:17589-17594.
48. KLEIN P, PAWSON T, TYERS M: Mathematical modeling suggests cooperative interactions between a disordered polyvalent ligand and a single receptor site. *Curr Biol.* (2003) **13**:1669-1678.
49. NASH P, TANG X, ORLICKY S, *et al.*: Multisite phosphorylation of a CDK inhibitor sets a threshold for the onset of DNA replication. *Nature* (2001) **414**:514-521.
50. KIRKWOOD TBL, KOWALD A: The network theory of aging. *Exp. Gerontol.* (1997) **32**:395-399.
51. ALBERT R, JEONG H, BARABASI A-L: Attack and error tolerance of complex networks. *Nature* (2000) **406**:378-382.
    • **A key paper on the vulnerability of networks on random and targeted attacks.**
52. COHEN R, EREZ K, BEN-AVRAHAM, D, HAVLIN, S: Resilience of the Internet to random breakdowns. *Phys. Rev. Lett.* (2000) **85**:4626-4628.
53. HOLME P, KIM BJ, YOON, CN, HAN, SK: Attack vulnerability of complex networks. *Phys. Rev. E* (2002) **65**:056109.
54. FREEMAN LC: A set of measuring centrality based on betweenness. *Sociometry* (1977) **40**:35-41.
55. AGOSTON V, CSERMELY P, PONGOR S: Multiple hits confuse complex systems: a genetic network as an example. *Phys. Rev. E* (2005) **71**:051909.
    • **An important early model of the potential effect of multi-target drugs on cellular networks.**
56. LATORA V, MARCHIORI, M: Efficient behavior of small-world networks. *Phys. Rev. Lett.* (2001) **87**:198701.
57. LATORA V, MARCHIORI, M: Vulnerability and protection of infrastructure networks. *Phys. Rev. E* (2005) **71**:015103.
58. DALL'ASTA L, BARRAT A, BARTHÉLEMY M, VESPIGNANI A. Vulnerability of weighted networks. *J. Stat. Mech.* (2006) P04006.
59. MOTTER AE, NISHIKAWA T, LAI Y: Range-based attack on links in scale-free networks: Are long-range links responsible for the small-world phenomenon? *Phys. Rev. E* (2002) **66**:065103.
60. KURANT M, THIRAN P: Error and attack tolerance of layered complex networks. (2006) www.arxiv.org/physics/0610018.
61. CASCANTE M, BOROS LG, COMIN-ANDUIX B, DE ATAURI P, CENTELLES JJ, LEE PWN: Metabolic control analysis in drug discovery and disease. *Nat. Biotechnol.* (2002) **20**:243-249.
62. CORNISH-BOWDEN A CARDENAS ML: Metabolic analysis in drug design. *C. R. Biol.* (2003) **326**:509-515.
63. DUARTE NC, BECKER SA, JAMSHIDI N *et al.*: Global reconstruction of the human metabolic network based on genomic and bibliomic data. *Proc. Natl. Acad. Sci. USA* (2007) **104**:1777-1782.
64. BERGMANN S, IHMELS J, BARKAI N: Similarities and differences in genome-wide expression data of six organisms. *PLoS Biology* (2004) **2**:85-93 (E9).
65. STUART JM, SEGAL E, KOLLER D, KIM SK: A gene-coexpression network for global discovery of conserved genetic modules. *Science* (2003) **302**:249-255.
66. ARCHAKOV AI, GOVORUN VM, DUBANOV AV *et al.*: Protein-protein interactions as a target for drugs in proteomics. *Proteomics* (2003) **3**:380-391.

67. AOYAMA T, SUZUKI Y, ICHIKAWA H: Neural networks applied to quantitative structure-activity relationship analysis. *J. Med. Chem.* (1990) **33**:2583-2590
68. CSERMELY P, AGOSTON V, PONGOR S: The efficiency of multi-target drugs: the network approach might help drug design. *Trends Pharmacol. Sci.* (2005) **26**:187-182.
    - •• **This paper first introduces the concept of network-based analysis in the action of multi-target drugs and lists several lines of evidence to show that low affinity, multi-target drugs may perturb networks more efficiently than high-affinity, single-hit drugs.**
69. HORNBERG JJ, BRUGGEMAN FJ, WESTERHOFF HV, LANKELMA J: Cancer: a Systems Biology disease. *Biosystems* (2006) **83**:81-90.
70. PALOP JJ, CHIN J, MUCKE L: A network dysfunction perspective on neurodegenerative diseases. *Nature* (2006) **443**:768-773.
71. CSERMELY P, SOTI C: Aging cellular networks: chaperones as major participants. *Exp. Gerontol.* (2007) **42**:113-119.
72. GARDNER TS, DI BERNARDO D, LORENZ D, COLLINS JJ: Inferring genetic networks and identifying compound mode of action via expression profiling. *Science* (2003) **301**:102-105.
    - • **An important example of dynamic network modelling of drug action.**
73. GIAEVER G, FLAHERTY P, KUMM J *et al.*: Chemogenomic profiling: identifying the functional interactions of small molecules in yeast. *Proc. Natl. Acad. Sci. USA* (2004) **101**:793-798.
74. LUM PY, ARMOUR CD, STEPANIATS SB, *et al.*: Discovering modes of action for therapeutic compounds using a genome-wide screen of yeast heterozygotes. *Cell* (2004) **116**:121-137.
75. PARSONS AB, LOPEZ A, GIVONI IE, et al.: Exploring the mode-of-action of bioactive compounds by chemical-genetic profiling in yeast. *Cell* (2006) **126**:611-625.
76. *IN VIVO* PHARMACOLOGY TRAINING GROUP: The fall and rise of *in vivo* pharmacology. *Trends Pharmacol.* (2002) **23**:13-18.
77. GOLDBERGER AL, AMARAL LAN, HAUSDORF JM, IVANOV PC, PENG C-K, STANLEY HE: Fractal dynamics in physiology: alterations with disease and aging. *Proc. Natl. Acad. Sci. USA* (2002) **99**:2466-2472.
78. SZALAY M, KOVACS IA, KORCSMAROS T, BODE C, CSERMELY P: Stress-induced rearrangements of cellular networks: consequences for protection and drug design. *FEBS Lett.* in press, arxiv.org/q-bio.MN/0702006.
79. HALLEN K, BJORKEGREN J, TEGNER J: Detection of compound mode of action by computational integration of whole-genome measurements and genetic perturbations. *BMC Bioinformatics* (2006) **7**:51.
    - •• **A careful analysis of network dynamics to assess drug action.**
80. KHOLODENKO BN, KIYATKIN A, BRUGGEMAN FJ, SONTAG E, WESTERHOFF HV, HOEK JB: Untangling the wires: A strategy to trace functional interactions in signaling and gene networks. *Proc. Natl. Acad. Sci. USA* (2002) **99**:12841-12846.
    - • **An important paper on the dynamical analysis of networks.**
81. TEGNÉR J, YEUNG MK, HASTY J, COLLINS JJ: Reverse engineering gene networks: integrating genetic perturbations with dynamical modeling. *Proc. Natl. Acad. Sci. USA* (2003) **100**:5944-5949.
    - • **An important paper on the dynamical analysis of networks.**

**Affiliation**
**Tamás Korcsmáros[1], Máté S Szalay[2], Csaba Böde[3] PhD, István A Kovács[4] & Péter Csermely[†5] PhD, DSc**
[†]Author for correspondence
[1]Department of Medical Chemistry, Semmelweis University, P O Box 260., H-1444 Budapest 8, Hungary and Predinet Ltd., Dongo str. 8., H-1149 Budapest, Hungary
[2]Department of Medical Chemistry, Semmelweis University, P O Box 260., H-1444 Budapest 8, Hungary and Predinet Ltd., Dongo str. 8., H-1149 Budapest, Hungary
[3]Department of Biophysics and Radiation Biology, Semmelweis University, P O Box 263., H-1444 Budapest 8, Hungary
[4]Department of Medical Chemistry, Semmelweis University, P O Box 260., H-1444 Budapest 8, Hungary and Predinet Ltd., Dongo str. 8., H-1149 Budapest, Hungary
[5]Department of Medical Chemistry, Semmelweis University, P O Box 260., H-1444 Budapest 8, Hungary and Predinet Ltd., Dongo str. 8., H-1149 Budapest, Hungary
Tel: +36 1 266 2755; Fax: +36 1 266 6550; E-mail: csermely@predinet.com

**Table 1. Cellular networks as drug target maps.**

| Name of cellular network | Network elements | Network links | Potential drug targets |
|---|---|---|---|
| Protein interaction network | Cellular proteins | Transient or permanent bonds | hubs, bridges, proteins in modular centres and overlaps |
| Cytoskeletal network | Cytoskeletal filaments | Transient or permanent bonds | cross-linking proteins |
| Organelle network | Membrane segments (membrane vesicles, domains, rafts, of cellular membranes) and cellular organelles (mitochondria, lysosomes, segments of the endoplasmic reticulum, etc.) | Proteins, protein complexes and/or membrane vesicles, channels | proteins and lipid rafts regulating inter-organellar junctions |
| Signalling network | Proteins, protein complexes, RNA (such as micro-RNA) | Highly specific interactions undergoing a profound change (either activation or inhibition), when a specific signal reaches the cell | hubs, bridge proteins of cross-talks, 'critical nodes' |
| Metabolic network | Metabolites, small molecules, such as glucose, or adenine, etc. | Enzyme reactions transforming one metabolite to the other | metabolic switch enzymes and their regulatory proteins, channelling |
| Gene transcription network | Transcriptional factors or their complexes and DNA gene sequences | Functional (and physical) interactions between transcription factor proteins (sometimes RNA-s) and various parts of the gene sequences in the cellular DNA | hubs, bridges, proteins in modular centres and overlaps |

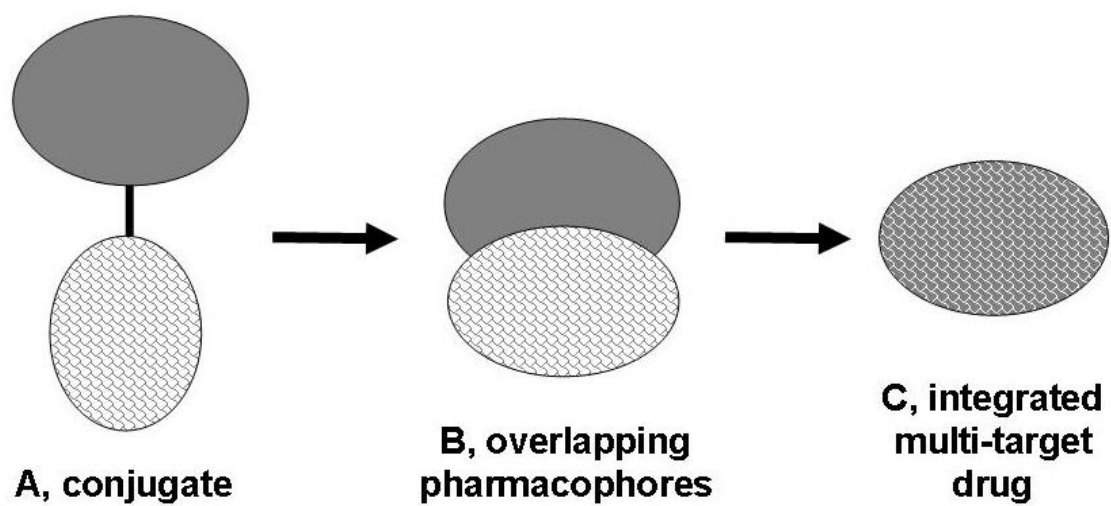

**Figure 1. From combinational therapy to multi-target drugs.** The increasing overlap of pharmacophores gives an almost continuous spectrum starting from conjugates (A) via slightly overlapping pharmacophores (B) till highly integrated multi-target drugs (C).

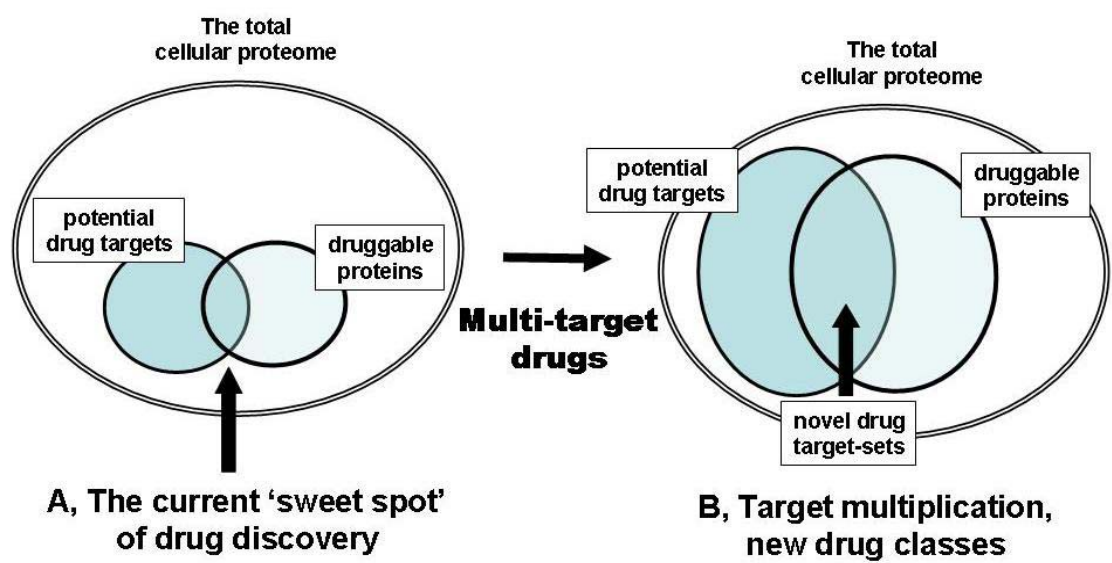

**Figure 2. From the 'sweet spot' of drug discovery to a potential candy-field.** Multi-target drugs may magnify the 'sweet spot' of drug discovery [Brown and Superti-Furga, 2003] to a whole candy-field. (A) The overlap between pathways, which are interesting from the pharmacological point of view and the hits of chemical proteomics, which represent those proteins which can interact with druggable molecules constitutes the 'sweet spot' of drug discovery. (B) Indirect effects of multi-target drugs expands the number of pharmacologically relevant targets, while low-affinity binding enlarges the number of druggable molecules.

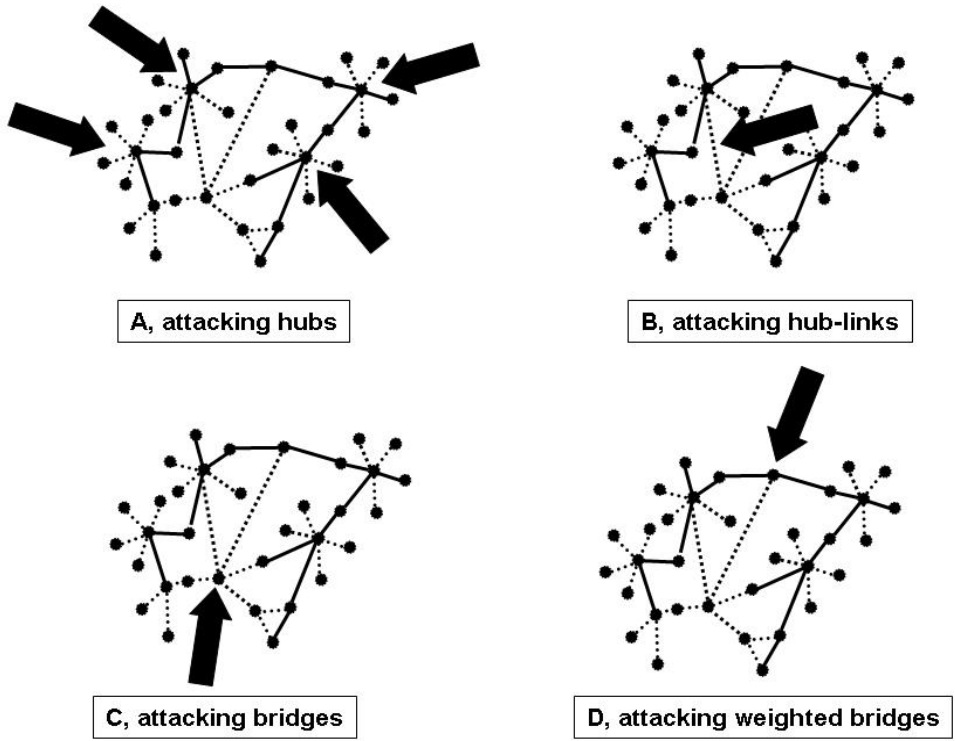

**Figure 3. Attack scenarios on networks.** In this figure we summarize a number of malicious attacks on vulnerable points of networks. (A) attacks on nodes with highest degree (hubs); (B) attacks on 'hub-links' with highest degree of their end-points; (C) attacks on bridging elements with links having a high betweenness centrality; (D) attacks on bridges having links with the highest weighted centrality.